\documentclass[12pt,a4paper]{article}
\usepackage{amsmath}
\usepackage{amssymb}
\usepackage{tikz}
\usepackage{cite}
\usepackage{hyperref}

\makeatletter
  \def\@seccntformat#1{%
    \@nameuse{@seccnt@prefix@#1}%
    \@nameuse{the#1}%
    \@nameuse{@seccnt@postfix@#1}%
    \@nameuse{@seccnt@afterskip@#1}}
  \def\@seccnt@prefix@section{}
  \def\@seccnt@postfix@section{.}
  \def\@seccnt@afterskip@section{\hspace{.5em}}
  \def\@seccnt@prefix@subsection{}
  \def\@seccnt@postfix@subsection{.}
  \def\@seccnt@afterskip@subsection{\hspace{.5em}}
\makeatother

\makeatletter
\renewcommand\section{
  \@startsection{section}{3}{\z@}%
  {-3.25ex\@plus -1ex \@minus -.2ex}%
  {1.5ex \@plus .2ex}%
  {\normalfont\normalsize\bfseries\mathversion{bold}}}
\renewcommand\subsection{
  \@startsection{subsection}{3}{\z@}%
  {-3.25ex\@plus -1ex \@minus -.2ex}%
  {1.5ex \@plus .2ex}%
  {\normalfont\normalsize\bfseries\mathversion{bold}}}
\makeatother

\makeatletter \@addtoreset{equation}{section} \makeatother
\renewcommand{\theequation}{\arabic{section}.\arabic{equation}}

\allowdisplaybreaks[1]

\let\oldthebibliography\thebibliography
\renewcommand\thebibliography[1]{
  \oldthebibliography{#1}\setlength{\itemsep}{0.4ex}}

\newcommand{\nn}{\nonumber}

\newcommand{\grp}[1]{\mathrm{#1}}
\newcommand{\bvec}[1]{\boldsymbol{#1}}
\newcommand{\tnum}[1]{\mbox{\footnotesize $#1$}}

\newcommand{\bbC}{\mathbb{C}}
\newcommand{\bbR}{\mathbb{R}}
\newcommand{\bbZ}{\mathbb{Z}}
\newcommand{\bbH}{\mathbb{H}}
\newcommand{\bbP}{\mathbb{P}}

\newcommand{\calA}{\mathcal{A}}
\newcommand{\calD}{\mathcal{D}}
\newcommand{\vecw}{{\boldsymbol{w}}}
\newcommand{\vecz}{{\boldsymbol{z}}}

\newcommand{\coeff}{\phi}
\newcommand{\tcoeff}{\tilde{\coeff}}
\newcommand{\gen}{\mathcal{E}}
\newcommand{\LK}{\Lambda_\mathrm{K3}}

\begin{document}


\def\papertitlepage{\baselineskip 3.5ex \thispagestyle{empty}}
\def\preprinumber#1#2{\hfill
\begin{minipage}{1.22in}
#1 \par\noindent #2
\end{minipage}}

%
\papertitlepage
\setcounter{page}{0}
\preprinumber{}{}
\vskip 2ex
\vfill
\begin{center}
{\large\bf\mathversion{bold}
F-theory on K3 surfaces and orthogonal modular forms
}
\end{center}
\vfill
\baselineskip=3.5ex
\begin{center}
Kazuhiro Sakai\\

{\small
\vskip 6ex
{\it Institute for Mathematical Informatics, Meiji Gakuin University\\
1518 Kamikurata-cho, Totsuka-ku,
Yokohama 244-8539, Japan}\\
\vskip 1ex
{\tt kzhrsakai@gmail.com}

}
\end{center}
\vfill
\baselineskip=3.5ex
\begin{center} {\bf Abstract} \end{center}

We identify the precise F-theory backgrounds
dual to the heterotic string theory compactified on $T^2$
with general Wilson lines turned on for one of the two $E_8$'s.
The elliptic curve describing the backgrounds
is expressed in terms of orthogonal modular forms
and is determined so that it admits a nontrivial scaling limit
yielding a rational elliptic surface.
We confirm that the curve reduces to
the Seiberg--Witten curve for the E-string theory in this limit.
We conjecture that our result gives an explicit inverse period map
for algebraic K3 surfaces polarized by
the $U\oplus E_8(-1)$ lattice.

\vfill
\noindent
September 2026


\setcounter{page}{0}
\newpage
\renewcommand{\thefootnote}{\arabic{footnote}}
\setcounter{footnote}{0}
\setcounter{section}{0}
\baselineskip = 3.5ex
\pagestyle{plain}

\section{Introduction}\label{sec:introduction}

Heterotic/F-theory duality
is a nontrivial equivalence of string theories
\cite{Vafa:1996xn,Morrison:1996na,Morrison:1996pp}. 
It has a close connection with algebraic geometry
and gives a rigorous foundation of string phenomenology.
While the equivalence continues to be extended to broader settings
in lower dimensions, the original duality in eight dimensions
is the source of all descendants and retains its enduring significance.

In this paper we provide a new refinement of this duality.
We consider the $E_8\times E_8$ heterotic string theory
compactified on $T^2$
with general Wilson lines turned on for one of the two $E_8$'s.
There are ten complex moduli:
the complex structure $\tau$ and
the complexified K\"ahler structure $\rho$ of $T^2$,
and eight parameters $z_1,\ldots,z_8$
that characterize the Wilson lines.
On the other hand, the system is dual to F-theory
compactified on an elliptic K3 surface of the form
\begin{align}
\begin{aligned}
y^2=4x^3
 -&\left(\coeff_4u^4+\coeff_{10}u^3+\coeff_{16}u^2
   +\coeff_{22}u+\coeff_{28}\right)x\\
 -&\left(-4u^7+\coeff_{12}u^5+\coeff_{18}u^4+\coeff_{24}u^3
   +\coeff_{30}u^2+\coeff_{36}u+\coeff_{42}\right).
\end{aligned}
\label{eq:O210curve}
\end{align}
The eleven coefficients $\coeff_k$
are nontrivially related to
the ten moduli $\tau,\rho,z_j$
on the heterotic side.
In this paper we obtain explicit expressions
for the functions $\coeff_k(\tau,\rho,z_j)$.

Mathematically, this is the problem of constructing 
an explicit inverse period map for lattice polarized K3 surfaces.
The Weierstrass model \eqref{eq:O210curve} describes
algebraic K3 surfaces polarized by the $U\oplus E_8(-1)$ lattice,
where $U$ is the even unimodular lattice of signature $(1,1)$.
There is a period map from the moduli space of
marked pseudo-ample K3 surfaces of the form \eqref{eq:O210curve}
to the period domain parametrized by $\tau,\rho,z_j$.
By Torelli-type theorem, an inverse map exists,
but its explicit form has not been known.
To date, explicit inverse period maps have been constructed
only for some limited families of K3 surfaces
\cite{Inose:1977,Kumar:2008,Clingher:2012,Malmendier:2014uka,
Nagano:2021}.\footnote{See also \cite{Clingher:2025,Clingher:2026}
for closely related recent work
discussing more general families of K3 surfaces
in relation to orthogonal modular forms.}

The theory of moduli space of K3 surfaces shows that
$\coeff_k(\tau,\rho,z_j)$ is
an orthogonal modular form of weight $k$
for the even unimodular lattice of signature $(2,10)$
\cite{Hashimoto:2022}.
Orthogonal modular forms are a generalization of
modular forms on $\grp{SL}(2,\bbZ)$.
It is known that the graded ring of orthogonal modular forms
for the above lattice
is a polynomial ring \cite{Hashimoto:2022,Looijenga:1984},
generated by eleven Eisenstein series \cite{Dieckmann:2019}.
This allows us to assume $\coeff_k$
as polynomials in these generators
with a finite number of undetermined coefficients.
Hence, the problem reduces to determining these coefficients.

We solve this problem using a simple principle.
We impose that \eqref{eq:O210curve}
admits a nontrivial scaling limit as $\rho\to i\infty$,
yielding a rational elliptic surface of the form
\begin{align}
\begin{aligned}
y^2=4x^3
 -&\left(\alpha_0u^4+\alpha_2u^2+\alpha_3u+\alpha_4\right)x\\
 -&\left(\beta_0u^6+\beta_1u^5+\beta_2u^4+\beta_3u^3+\beta_4u^2
  +\beta_5u+\beta_6\right).
\end{aligned}
\label{eq:Ecurve}
\end{align}
More specifically, 
as we will see shortly, it is reasonable
to expect that $\alpha_m(\tau,z_j)$, $\beta_m(\tau,z_j)$
are (meromorphic) Weyl invariant $E_8$ Jacobi forms of index $m$.
On the other hand,
every orthogonal modular form $\phi_k(\tau,\rho,z_j)$ has
a Weyl invariant $E_8$ Jacobi form of index $m$ as
the coefficient of $e^{2\pi i m\rho}$
in its Fourier--Jacobi expansion.
Thus, we impose conditions under which
appropriate Fourier--Jacobi coefficients
are singled out in the scaling limit.
Remarkably, this uniquely determines the functions $\coeff_k$.

The above scaling limit is physically interpreted as follows.
The F-theory on K3 describes
the type IIB string theory in $\bbR^8\times\bbP^1$
with 24 7-branes filling the $\bbR^8$.
The $\bbP^1$ is identified with the base space of the elliptic K3.
If we view \eqref{eq:O210curve} as an elliptic curve,
its discriminant is a polynomial in $u$ of degree 14,
whose zeros give the locations of 14 7-branes
in the region $u<\infty$ of the $\bbP^1$.
We can select 12 7-branes whose total monodromy is trivial.
The scaling limit corresponds to
sending the other two 7-branes to $u=\infty$
while keeping the twelve in the region $u<\infty$.
If we probe the resulting background
by a single D3-brane,
the low-energy theory is that of the E-string theory
\cite{Ganor:1996mu,Seiberg:1996vs} in $\bbR^4\times T^2$
\cite{Ganor:1996pc}.
The background is described by
the Seiberg--Witten curve for the E-string theory \cite{Eguchi:2002fc},
which is exactly of the form \eqref{eq:Ecurve},
with the coefficients $\alpha_m,\beta_m$
being meromorphic Weyl invariant $E_8$ Jacobi forms.

In accordance with the physical expectation, we confirm that
our solution indeed reduces to
the Seiberg--Witten curve for the E-string theory in the scaling limit.
We emphasize that the only assumptions on the curve \eqref{eq:Ecurve}
we make are that $\alpha_m,\beta_m$ are Jacobi forms of index $m$
and that $\alpha_0,\beta_0$ satisfy appropriate normalization
conditions; no further conditions are imposed.
Therefore, we not only obtain the F-theory backgrounds,
but also rederive
the Seiberg--Witten curve for the E-string theory.
The curve was originally derived in \cite{Eguchi:2002fc}
by using mirror symmetry 
for certain Calabi--Yau threefolds \cite{Minahan:1998vr}.
Our present approach gives a much simpler derivation.

The emergence of the E-string curve provides strong evidence that
we obtain the correct F-theory duals, since the curve
accurately translates symmetry breaking patterns
induced by Wilson lines into splitting patterns of singular fibers
\cite{Eguchi:2002nx,Sakai:2012ik,Sakai:2017ihc}.
Mathematically, we conjecture that
our result gives an explicit inverse period map
for algebraic K3 surfaces polarized by
the $U\oplus E_8(-1)$ lattice.
We test this conjecture by adjusting the Wilson line parameters
$z_j$ to special values. We see that our result reproduces
the known explicit inverse period map
\cite{Kumar:2008,Clingher:2012,Malmendier:2014uka}
for algebraic K3 surfaces polarized by
the $U\oplus E_8(-1)\oplus E_7(-1)$ lattice,
described in terms of Siegel modular forms of genus two.

The paper is organized as follows.
In Section~\ref{sec:duality}, we review the heterotic/F-theory
duality in eight dimensions
and the period map for lattice-polarized K3 surfaces.
In Section~\ref{sec:Jacobi}, we recall
some useful facts about Weyl-invariant $E_8$ Jacobi forms.
In Section~\ref{sec:ortho},
we review orthogonal modular forms
and recall explicit expressions for the Eisenstein series.
In Section~\ref{sec:sol},
we construct our solution and discuss its properties.
In Section~\ref{sec:red},
we demonstrate that our solution reduces to known ones.
We conclude in Section~\ref{sec:conclusion}.
Explicit expressions for $\coeff_k$ are presented
in Appendix~\ref{app:coeff}.

\section{Heterotic/F-theory duality and period map for K3 surfaces}
\label{sec:duality}

In this section we review the heterotic/F-theory duality
in eight dimensions \cite{Vafa:1996xn,Morrison:1996na,Morrison:1996pp}
and the period map for lattice-polarized K3 surfaces.

F-theory describes nontrivial vacua of type IIB string theory.
In this paper we consider F-theory compactified on
elliptic K3 surfaces with section.
It describes IIB string theory
in $\bbR^8\times\bbP^1$ with certain nontrivial backgrounds.
The $\bbP^1$ is identified with the base space of the elliptic K3.
Typically, the background fields vary over the $\bbP^1$.
In particular, the background value of the axio-dilaton field
is given by the complex structure modulus $\tilde{\tau}(u)$
of the elliptic fiber, where $u$ is the coordinate of $\bbP^1$.
The other fields are also described
in terms of the elliptic fibration.
The axio-dilaton background $\tilde{\tau}(u)$
has singularities in the $\bbP^1$.
This is allowed, indicating that 7-branes sit at the singularities
and extend along the $\bbR^8$.
The singularities are the points where the elliptic fiber
degenerates, i.e.~the zeros of the discriminant of the fiber.
For a generic elliptic K3,
the discriminant has 24 zeros (counting multiplicities),
which give the locations of 24 7-branes.
Multiple zeros correspond to coincident 7-branes.

The heterotic/F-theory duality in eight dimensions is
the equivalence between
the $E_8\times E_8$ (or $\grp{SO}(32)$) heterotic string theory
compactified on $T^2$ and
F-theory compactified on elliptic K3 surfaces.
On the heterotic side, there are 18 complex moduli:
the complex structure $\tau$ and
the complexified K\"ahler structure $\rho$ of $T^2$,
and 16 complex parameters that characterize the Wilson lines for the
$E_8\times E_8$ gauge group around $T^2$.
The number of complex moduli exactly matches that of
elliptic K3 surfaces \cite{Vafa:1996xn,Morrison:1996pp}.
In addition, there is a real positive coupling constant
on the heterotic side, which is identified with the size of
the $\bbP^1$ on the F-theory side.
Moreover, at least for small heterotic coupling, 
there has been established
a precise identification of the moduli spaces:
The moduli space of the heterotic string theory
compactified on $T^2$ is given by
$\Gamma\backslash\grp{O}(2,18)/(\grp{SO}(2)\times\grp{O}(18))$
\cite{Narain:1985jj,Narain:1986am},
where $\Gamma$ is the group of integral isometries of
the underlying lattice.
On the F-theory side, the same space arises
as the coarse moduli space of elliptic K3 surfaces with section.
See \cite{Clingher:2003ui} for
a mathematically rigorous treatment.
In this paper we rely on this identification.

We are interested in the explicit duality map.
To see this, let us start with a simple situation.
It is well known that
the $E_8\times E_8$ heterotic string theory
compactified on $T^2$ without Wilson lines
is dual to F-theory compactified on an elliptic K3 surface of the form
\cite{Morrison:1996pp}
\begin{align}
y^2=x^3+\alpha u^4x+u^7+\beta u^6+u^5.
\label{eq:mlcurve1}
\end{align}
Here $\alpha$ and $\beta$ are certain functions of $\tau$ and $\rho$
whose explicit forms are known \cite{Inose:1977}.
This K3 surface admits two singular fibers of 
Kodaira type $\mathrm{II}^\ast$, each at $u=0$ and $u=\infty$.
These reflect the unbroken $E_8\times E_8$ symmetry
on the heterotic side.

If we turn on general Wilson lines for one of the two $E_8$'s,
the K3 surface \eqref{eq:mlcurve1} is deformed partially so that
one of the two singular fibers, say the one at $u=\infty$,
is kept intact.
The surface then takes the form \eqref{eq:O210curve},
where we have rescaled $y$ by $y\mapsto y/2$
and translated $u$ to eliminate the $u^6$ term.
The subscript $k$ of $\coeff_k$ indicates its weighted degree,
where $u,x,y$ have weights $6,14,21$, respectively,
and the equation has total weight $42$.
The surface \eqref{eq:O210curve} admits equivalence given 
by the simultaneous rescaling $\coeff_k\mapsto\lambda^k\coeff_k$
for any $\lambda\in\bbC^\ast$,
since this is compensated by the change of coordinates
$(u,x,y)\mapsto(\lambda^6u,\lambda^{14}x,\lambda^{21}y)$.
Hence, the actual moduli space is a ten-dimensional
space of the projective classes
$[\phi_4:\phi_{10}:\cdots:\phi_{42}]$.
To be more precise, we have to exclude the surfaces
that can be transformed into the form 
\begin{align}
y^2=4x^3+au^4x+4u^7+bu^6
\label{eq:singular}
\end{align}
by a translation of $u$,
since they have a singularity worse than rational double points 
and are no longer K3 surfaces.
See \cite{Hashimoto:2022} for a more detailed and mathematically
rigorous description of the moduli space.

The surfaces of the form \eqref{eq:O210curve}
are known as $(U\oplus E_8(-1))$-polarized K3 surfaces.
We briefly recall the notion of
lattice-polarized K3 surfaces \cite{Nikulin:1979,Dolgachev:1996xw}.
Let $X$ be a complex algebraic K3 surface.
The second cohomology group $H^2(X,\bbZ)$
is endowed with a lattice structure via the cup-product.
The lattice is called the K3 lattice and has the structure
\begin{align}
\LK=U\oplus U\oplus U\oplus E_8(-1)\oplus E_8(-1).
\end{align}
Here, $U$ is the even unimodular lattice of signature $(1,1)$
and $E_8(-1)$ denotes
the even unimodular negative definite lattice of rank $8$.
Let $S$ be an even non-degenerate lattice of signature $(1,t)$
embeddable into $\LK$.
An $S$-polarized K3 surface is a pair $(X,j)$ of a K3 surface $X$
and a primitive lattice embedding $j:S\hookrightarrow\mathrm{Pic}(X)$.
Here, $\mathrm{Pic}(X)$ denotes the Picard group of $X$.
A marked $S$-polarized K3 surface is a pair $(X,\phi)$
of a K3 surface $X$
and a lattice isomorphism $\phi:H^2(X,\bbZ)\to \LK$
such that $\phi^{-1}(S)\subset\mathrm{Pic}(X)$.

We also recall some essential facts about the period map.
Given a marked $S$-polarized K3 surface $(X,\phi)$,
the Hodge decomposition of $H^2(X,\bbC)$
defines the period point $\phi(H^{2,0}(X))$ in
$\bbP(\LK\otimes\bbC)$.
As explained in \cite{Dolgachev:1996xw},
the period point $\phi(H^{2,0}(X))$ in fact lies in the subspace
\begin{align}
\calD_S=
\{[Z]\in\bbP(S^\perp\otimes\bbC):(Z,Z)=0,\,(Z,\bar{Z})>0\}.
\label{eq:calDS}
\end{align}
Here, $S^\perp$ is the orthogonal complement of $S$ in $\LK$.
This space consists of two connected components $\calD^\pm_S$,
each isomorphic to the bounded Hermitian symmetric domain
of type $\mathrm{IV}_{19-t}$.
The above map induces
a holomorphic period map from the fine moduli space of
marked pseudo-ample $S$-polarized K3 surfaces
to the period domain $\calD_S^+$.
The global Torelli theorem \cite{Pyatetski-Shapiro:1971,Burns:1975}
and surjectivity theorem \cite{Todorov:1980}
show that this map is an isomorphism.
Put differently,
the coarse moduli space of pseudo-ample $S$-polarized K3 surfaces
is given by $\Gamma\backslash\calD_S^+$,
where $\Gamma=\grp{O}^+(S^\perp)$ is the index-two subgroup of
the integral isometries of $S^\perp$ that preserves $\calD_S^+$.
Note that the restriction to pseudo-ample surfaces
ensures the injectivity of the period map.

Let us now come back to the duality.
It is well known that the space \eqref{eq:calDS}
can be identified with the homogeneous space
\begin{align}
\frac{\grp{O}(2,19-t)}{\grp{SO}(2)\times\grp{O}(19-t)}.
\end{align}
General elliptic K3 surfaces with section
are $U$-polarized K3 surfaces.
In this case, we have $t=1$ and the coarse moduli space is given by
$\Gamma\backslash\grp{O}(2,18)/(\grp{SO}(2)\times\grp{O}(18))$,
as mentioned earlier.
In this paper we consider the family \eqref{eq:O210curve},
which corresponds to the choice $S=U\oplus E_8(-1)$.
As shown in \cite{Hashimoto:2022},
apart from the singular exceptions leading to \eqref{eq:singular},
any elliptic K3 of the form \eqref{eq:O210curve}
has a natural structure of a pseudo-ample
$S$-polarized K3 surface.
We have $S^\perp=U\oplus U\oplus E_8(-1)$
and the corresponding period domain $\calD_S^+$
is the bounded Hermitian symmetric domain of type $\mathrm{IV}_{10}$.
As explained above,
the duality map is viewed as an inverse period map,
which respects the integral isometries of $S^\perp$
and hence is described naturally by modular forms
for this isometry group. We will study these modular forms
in Section~\ref{sec:ortho}.

\section{Weyl invariant $E_8$ Jacobi forms}\label{sec:Jacobi}

Jacobi forms appear in the Fourier--Jacobi expansion of orthogonal
modular forms. In this section we recall
the definition of Weyl invariant $E_8$ Jacobi forms
and some useful facts.
See \cite{Wang:2018fil,Sakai:2024vby} for further details.

Let $\bbH=\{\tau\in\bbC:\mathrm{Im}\,\tau>0\}$
be the upper half-plane and set $q=e^{2\pi i\tau}$.
Let $W(E_8)$ denote the Weyl group of the $E_8$ root system.
The $E_8$ lattice is also simply expressed as $E_8$.
A $W(E_8)$-invariant weak Jacobi form of weight $k$
and index $m$ ($k\in\bbZ,\ m\in\bbZ_{\ge 0}$) 
is a holomorphic function 
$\varphi:\bbH\times(E_8\otimes\bbC)\to\bbC$
that satisfies the following conditions
\cite{Eichler:1985, Wirthmuller:1992}:
\renewcommand{\theenumi}{\roman{enumi}}
\renewcommand{\labelenumi}{(\theenumi)}
\begin{enumerate}
\item Weyl invariance:
\begin{align}
\varphi(\tau,w(\vecz))=\varphi(\tau,\vecz),
\qquad w\in W(E_8).
\end{align}

\item Quasi-periodicity:
\begin{align}
\varphi(\tau,\vecz+\tau\bvec{\alpha}+\bvec{\beta})
=e^{-m \pi i (\tau\bvec{\alpha}^2+2\vecz\cdot\bvec{\alpha})}
\varphi(\tau,\vecz),\qquad
\bvec{\alpha},\bvec{\beta}\in E_8.
\end{align}

\item Modular transformation law:
\begin{align}
&
\varphi\left(
\frac{a\tau+b}{c\tau+d}\,,\frac{\vecz}{c\tau+d}\right)
=(c\tau+d)^k\exp\left(m\pi i\frac{c}{c\tau+d}\,\vecz^2\right)
\varphi(\tau,\vecz),\\[1ex]
&
\begin{pmatrix}a&b\\ c&d\end{pmatrix} \in \grp{SL}(2,\bbZ).\nn
\end{align}

\item It admits a Fourier expansion of the form
\begin{align}
\varphi(\tau,\vecz)
=\sum_{n=0}^\infty\sum_{\vecw\in E_8}
 c(n,\vecw)e^{2\pi i\vecw\cdot\vecz}q^n.
\label{eq:Fourier}
\end{align}
\end{enumerate}
If $\varphi(\tau,\vecz)$ further satisfies the condition
that the coefficients $c(n,\vecw)$
of the Fourier expansion \eqref{eq:Fourier}
vanish unless $\vecw^2\le 2mn$,
it is called a $W(E_8)$-invariant holomorphic Jacobi form.
If we say that $\psi(\tau,\vecz)$ is
a $W(E_8)$-invariant meromorphic Jacobi form, we mean that 
$\psi$ itself is not a $W(E_8)$-invariant weak Jacobi form
but there exists a modular form $f(\tau)$ on $\grp{SL}(2,\bbZ)$
such that $f\psi$ is a $W(E_8)$-invariant weak Jacobi form.

A modular form on $\grp{SL}(2,\bbZ)$
is a $W(E_8)$-invariant holomorphic Jacobi form
of index $0$, and vice versa.
The theta function associated with the $E_8$ lattice
\begin{align}
\begin{aligned}
\Theta(\tau,\vecz)
 =\sum_{\vecw\in E_8}
   \exp\left(\pi i\tau\vecw^2
   +2\pi i\vecz\cdot\vecw\right)
 =\frac{1}{2}\sum_{k=1}^4\prod_{j=1}^8\vartheta_k(z_j,\tau)
\end{aligned}
\label{eq:Theta}
\end{align}
is a $W(E_8)$-invariant holomorphic Jacobi form
of weight $4$ and index $1$.
Here, $\vartheta_k(z,\tau)$ are the Jacobi theta functions.

It is convenient to express $W(E_8)$-invariant Jacobi forms
as polynomials in some fundamental building blocks.
Unlike the cases of the other irreducible root systems
\cite{Wirthmuller:1992},
the bigraded ring of $W(E_8)$-invariant weak Jacobi forms is
not a simple polynomial ring \cite{Wang:2018fil}.
The ring was recently identified with
a certain joint covariant ring of binary forms
and a minimal basis of generators was obtained \cite{Sakai:2024vby}.
For our purposes, however, it suffices to know the following fact:
any $W(E_8)$-invariant weak Jacobi form
can be written uniquely as a polynomial in
11 algebraically independent
functions $a_0,a_2,a_3,a_4,b_0,b_1,\ldots,b_6$
\cite{Sakai:2022taq}.
More specifically,
\begin{align}
a_0=\frac{E_4(\tau)}{12},\qquad b_0=\frac{E_6(\tau)}{216}
\label{eq:a0b0}
\end{align}
are the Eisenstein series for $\grp{SL}(2,\bbZ)$
and the other $a_i(\tau,\vecz),b_j(\tau,\vecz)$
are $W(E_8)$-invariant meromorphic Jacobi forms.
In fact, these functions are the coefficients
of the Seiberg--Witten curve for the E-string theory
obtained in \cite{Eguchi:2002fc,Sakai:2011xg}.
They play a key role in the study of
$W(E_8)$-invariant weak Jacobi forms \cite{Sakai:2024vby}.

\section{Orthogonal modular forms}\label{sec:ortho}

In this section we summarize some basic facts about
orthogonal modular forms.
See e.g.~\cite{Gritsenko:2018} for a more detailed discussion.

Let $L$ be the even unimodular lattice of signature $(2,10)$
and $(\cdot,\cdot)$ denote its bilinear form.
We extend the bilinear form to the complexification of $L$.
The lattice is decomposed as $L=U\oplus U\oplus E_8(-1)$,
which actually appeared in Section~\ref{sec:duality}
as the orthogonal complement $S^\perp$
of $S=U\oplus E_8(-1)$ in $\LK$.
The bounded Hermitian symmetric domain of type $\mathrm{IV}_{10}$
is given by
\begin{align}
\calD=
\{[Z]\in\bbP(L\otimes\bbC) : (Z,Z)=0,\ (Z,\bar{Z})>0\}^+,
\end{align}
where the superscript ${}^+$ indicates
a choice of one connected component.
Let $\grp{O}(L)$ be the group of integral isometries of $L$
and $\grp{O}^+(L)$ denote the subgroup of $\grp{O}(L)$
that preserves $\calD$.
Let $\calA=\{Z\in L\otimes\bbC\setminus\{0\}:[Z]\in\calD\}$
be the affine cone over $\calD$.
A modular form of weight $k\in\bbZ_{\ge 0}$
with respect to the group $\grp{O}^+(L)$ is a holomorphic function
$F:\calA\to\bbC$
that satisfies the following conditions:
\renewcommand{\theenumi}{\roman{enumi}}
\renewcommand{\labelenumi}{(\theenumi)}
\begin{enumerate}
\item $F(\lambda Z)=\lambda^{-k}F(Z)$ for all $\lambda\in\bbC^\ast$,
\item $F(g Z)=F(Z)$ for all $g\in\grp{O}^+(L)$.
\end{enumerate}
We call it simply an orthogonal modular form for $L$.

Orthogonal modular forms can be efficiently
studied via their Fourier--Jacobi expansions.
Recall the decomposition $L=U\oplus U\oplus E_8(-1)$
and let $e_i,f_i\ (i=1,2)$ be primitive isotropic vectors
in the two $U$-lattices:
$(e_i,e_i)=(f_i,f_i)=0,\ (e_i,f_i)=1$.
Then any point in $\calD$ can be uniquely expressed as
the projective class $[Z]$ of
\begin{align}
Z=\bigl(\tfrac{1}{2}\vecz^2-\tau\rho\bigr)e_1+f_1
 +\tau e_2+\rho f_2+\vecz,
\end{align}
where $\tau,\rho\in\bbH,\ \vecz=(z_1,\ldots,z_8)\in E_8\otimes\bbC$
and $2(\mathrm{Im}\tau)(\mathrm{Im}\rho)>\sum_j(\mathrm{Im}z_j)^2$. 
With this parametrization,
a modular form $F$ is viewed as a function of ten variables:
\begin{align}
\phi(\tau,\rho,\vecz)
 :=F\bigl((\tfrac{1}{2}\vecz^2-\tau\rho)e_1+f_1
   +\tau e_2+\rho f_2+\vecz\bigr).
\end{align}
It admits a Fourier--Jacobi expansion
\begin{align}
\phi(\tau,\rho,\vecz)
 =\sum_{m=0}^\infty\varphi_m(\tau,\vecz)p^m,\qquad
p=e^{2\pi i\rho}.
\label{eq:generalFJ}
\end{align}
A key fact is that the coefficient $\varphi_m$ is
a $W(E_8)$-invariant holomorphic Jacobi form
of weight $k$ and index $m$, where $k$ is the weight of $\phi$.
Another fundamental property of $\phi$ is that it is
symmetric in $\tau$ and $\rho$:
\begin{align}
\phi(\rho,\tau,\vecz)=\phi(\tau,\rho,\vecz).
\end{align}

Orthogonal modular forms form a graded ring.
It is known \cite{Hashimoto:2022,Looijenga:1984} that
the graded ring of modular forms for $L$ is a polynomial ring
generated by forms of weight $k\in K$ with
\begin{align}
K=\{4,10,12,16,18,22,24,28,30,36,42\}.
\end{align}
One can choose the Eisenstein series $\gen_k$ of weight $k\in K$
as generators \cite{Dieckmann:2019}.
Each $\gen_k$ can be constructed as the Maass lift of
a $W(E_8)$-invariant holomorphic Jacobi form
of weight $k$ and index $1$:
its Fourier--Jacobi expansion is written
explicitly as \cite{Eie:1991}
\begin{align}
\gen_k(\tau,\rho,\vecz)
 =-\frac{B_k}{2k}E_k(\tau)
  +\sum_{m=1}^\infty
   \left(E_{k-4}(\tau)\Theta(\tau,\vecz)\right)\big|_{T(m)}p^m.
\label{eq:genFJ}
\end{align}
Here, $B_k$ is the $k$th Bernoulli number,
$E_k(\tau)$ is the normalized Eisenstein series of weight $k$
for $\grp{SL}(2,\bbZ)$ (we set $E_0(\tau)=1$),
and $\Theta(\tau,\vecz)$ is the theta function \eqref{eq:Theta}.
The operation of the Hecke operator $T(m)$ on $\varphi(\tau,\vecz)$
of weight $k$ is written as
\begin{align}
\varphi\big|_{T(m)}(\tau,\vecz)
 =m^{k-1}\sum_{\substack{ad=m,\; a\ge 1\\ 0\le b<d}}
  d^{-k}\varphi\left(\frac{a\tau+b}{d},a\vecz\right).
\end{align}
%

\section{Construction of the solution}\label{sec:sol}

In this section we construct the F-theory duals
by determining the coefficient functions $\coeff_k(\tau,\rho,\vecz)$ of
the elliptic K3 surface \eqref{eq:O210curve}.\footnote{We often
refer to \eqref{eq:O210curve} as an elliptic curve,
with $u$ understood as a parameter.}
We also discuss some notable properties of the solution.

As we saw in Section~\ref{sec:introduction}, the brane probe picture
implies that the elliptic curve describing the F-theory backgrounds
encodes all the information about
the moduli space of the E-string theory.
It is therefore reasonable to assume that the curve reduces to
the Seiberg--Witten curve for the E-string theory
in an appropriate limit. Mathematically, this induces
a degeneration of the elliptic K3 surface \eqref{eq:O210curve}
to a rational elliptic surface of the form \eqref{eq:Ecurve}.
It is unlikely that such a degeneration
could occur at a generic period point.
It should occur somewhere at the boundary of the period domain.
Since the E-string curve is independent of $\rho$,
it is reasonable to expect the degeneration
to occur at the cusp $\rho=i\infty$.
We thus consider the limit $p\to 0$.
As the E-string curve has $W(E_8)$-invariant
Jacobi forms of higher index as its coefficients,
we formulate a scaling limit such that
appropriate Fourier--Jacobi coefficients of orthogonal modular forms
are singled out as $p\to 0$.

We first express the Fourier--Jacobi expansions of
the 11 generators $\gen_k(\tau,\rho,\vecz)$
in terms of the 11 $W(E_8)$-invariant meromorphic Jacobi forms
$a_i(\tau,\vecz),b_j(\tau,\vecz)$ given in \cite{Sakai:2011xg}.
This is a key technical idea behind our construction,
making subsequent computations transparent.
We use the explicit formula \eqref{eq:genFJ},
where Fourier--Jacobi coefficients are
by construction
$W(E_8)$-invariant holomorphic Jacobi forms.
As mentioned in Section~\ref{sec:Jacobi}, any $W(E_8)$-invariant
Jacobi form can be expressed uniquely as a polynomial in $a_i,b_j$.
For our purposes, it suffices to compute these expressions
for each $\gen_k$ up to and including the $p^6$ term.
The computation is technically involved
but conceptually straightforward.
We obtain, for instance,\footnote{Explicit
Fourier--Jacobi expansions of $\gen_k$
up to and including the $p^6$ term
are available in the ancillary file \texttt{FJ.txt}.}
\begin{align}
\begin{aligned}
\gen_4&=\frac{1}{20}a_0-3a_0b_1p
 +\left(-\frac{3}{2}a_0^3a_2+\frac{27}{4}a_0b_1^2
  +\frac{81}{2}a_2b_0^2\right)p^2+O(p^3),\\
\gen_{10}&=-\frac{108}{11}a_0b_0-648a_0b_0b_1p
  +\bigl(4536a_0^4b_2-7128a_0^3a_2b_0-122472a_0b_0^2b_2\\
&\hspace{13em}
  +83106a_0b_0b_1^2+192456a_2b_0^3\bigr)p^2+O(p^3).
\end{aligned}
\label{eq:FJgen}
\end{align}

We then move on to the curve \eqref{eq:O210curve}.
As Hashimoto and Ueda identified in \cite{Hashimoto:2022},
each coefficient function $\coeff_k(\tau,\rho,\vecz)$
is regarded as an orthogonal modular form of weight $k$ for $L$.
This means that $\phi_k$ can be written as a polynomial
in the 11 Eisenstein series $\gen_l$.
Let us make the most general ansatz subject to the grading.
This involves 76 undetermined coefficients $c_1,\ldots,c_{76}$:
\begin{align}
\begin{aligned}
\coeff_4&=c_1\gen_4,\quad
\coeff_{10}=c_2\gen_{10},\quad
\coeff_{12}=c_3\gen_{12}+c_4\gen_4^3,\quad
\ldots,\quad\\
\coeff_{42}&=c_{54}\gen_{42}+\cdots+c_{76}\gen_4^8\gen_{10}.
\end{aligned}
\end{align}

Next, to bring the curve \eqref{eq:O210curve}
into a form similar to \eqref{eq:Ecurve},
we make the following change of variable
\begin{align}
u\mapsto u-\frac{\coeff_{10}}{4\coeff_4}.
\label{eq:uturel}
\end{align}
This eliminates the $u^3x$ term and restores the $u^6$ term.
The curve is rewritten as
\begin{align}
\begin{aligned}
y^2=4x^3
 -&\left(\tcoeff_4u^4
   +\tcoeff_{16}u^2+\tcoeff_{22}u+\tcoeff_{28}\right)x\\
 -&\left(-4u^7+\tcoeff_6u^6+\tcoeff_{12}u^5+\tcoeff_{18}u^4
   +\tcoeff_{24}u^3+\tcoeff_{30}u^2+\tcoeff_{36}u
   +\tcoeff_{42}\right),
\end{aligned}
\label{eq:tO210curve}
\end{align}
where $\tcoeff_k$ are
certain rational expressions in $\coeff_l$.

After that, we require
$\tcoeff_k$ to have the following Fourier--Jacobi expansions:
\begin{align}
\begin{aligned}
\tcoeff_4&=a_0+O(p),\qquad
\tcoeff_{4+6i}=O(p^i),\quad i=2,3,4,\\
\tcoeff_6&=b_0+O(p),\qquad
\tcoeff_{6+6j}=O(p^j),\quad j=1,2,3,4,5,6.
\end{aligned}
\label{eq:FJcond}
\end{align}
Here, $a_0$ and $b_0$ are the Eisenstein series \eqref{eq:a0b0}.
The conditions on $\tcoeff_4$ and $\tcoeff_6$
fix the rescaling gauge freedom $\coeff_k\mapsto\lambda^k\coeff_k$.
They also encode the `boundary condition' that
the elliptic fiber at $u=\infty$ after the scaling
is of complex modulus $\tau$.
We will see that the $u^7$ term vanishes in the scaling limit
and hence the $u^4x$ and $u^6$ terms become dominant
as $u\to\infty$.
Note that originally we have a singular fiber of
type $\mathrm{II}^\ast$ at $u=\infty$,
for which the complex modulus is fixed at $e^{2\pi i/3}$.
After the scaling we generically have a regular fiber at $u=\infty$,
which can be of any complex modulus $\tau\in\bbH$.
The conditions on the other $\tcoeff_k$'s naturally single out
$W(E_8)$-invariant Jacobi forms of higher index as $p\to 0$.
By using the Fourier--Jacobi expansion formulas \eqref{eq:FJgen}
and the algebraic independence of $a_i,b_j$,
the conditions \eqref{eq:FJcond} yield
an overdetermined linear system of equations
for the undetermined coefficients $c_1,\ldots,c_{76}$.
Remarkably, the equations uniquely determine all $c_i$'s.
We finally obtain
\begin{align}
\begin{aligned}
\coeff_4&=
  20\gen_4,\qquad
\coeff_{10}=
\hspace{-.2em}{}
 -\frac{11}{756}\gen_{10},\qquad
\coeff_{12}=
  \frac{13}{21600}\gen_{12}
 -56\gen_4^3,\\[1ex]
\coeff_{16}&=
  \frac{11747}{24766560000}\gen_{16}
 -\frac{151}{54000}\gen_4\gen_{12}
 +\frac{1564}{15}\gen_4^4,\qquad\ldots.
\end{aligned}
\end{align}
The full results are presented in Appendix~\ref{app:coeff}.

Given the solution, we can compute the leading terms
of the Fourier--Jacobi expansions of $\tcoeff_k$.
The results are surprisingly simple
when expressed in terms of $a_i,b_j$.
We obtain
\begin{align}
\tcoeff_{4+6n}=a_n \Delta^n p^n +O(p^{n+1}),\qquad
\tcoeff_{6+6n}=b_n \Delta^n p^n +O(p^{n+1}),
\label{eq:tcoeffexp}
\end{align}
where $\Delta=\eta(\tau)^{24}=a_0^3-27b_0^2$ and $\eta(\tau)$
is the Dedekind eta function.
These results imply the following:
By rescaling the variables as
\begin{align}
u\mapsto(\Delta p)u,\quad
x\mapsto(\Delta p)^2x,\quad
y\mapsto(\Delta p)^3y,
\label{eq:Dprescaling}
\end{align}
equation \eqref{eq:tO210curve} becomes
\begin{align}
\begin{aligned}
y^2=4x^3
 -&\left(a_0u^4+a_2u^2+a_3u+a_4\right)x\\
 -&\left(b_0u^6+b_1u^5+b_2u^4
   +b_3u^3+b_4u^2+b_5u
   +b_6\right)
+O(p).
\end{aligned}
\label{eq:SWcurve1}
\end{align}
The $u^7$ term is now relegated to the $O(p)$ part.
In the limit $p\to 0$,
the above equation exactly reproduces the Seiberg--Witten curve
for the E-string theory \cite{Eguchi:2002fc}.

The emergence of the E-string curve provides strong evidence that
we obtain the correct F-theory duals,
since the curve is known to
accurately translate symmetry breaking patterns
induced by Wilson lines into splitting patterns of singular fibers
\cite{Eguchi:2002nx,Sakai:2012ik,Sakai:2017ihc}.
This is the very property that the F-theory solution
and the corresponding inverse period map must have.
We therefore conjecture that
our result gives an explicit inverse period map
for $(U\oplus E_8(-1))$-polarized K3 surfaces.
We leave the proof as an open problem for future work.
Instead, in the next section
we will show that the conjecture passes a nontrivial test.

We close this section with a few remarks.
First, if we set $p=0$ without applying
the rescaling \eqref{eq:Dprescaling},
equation \eqref{eq:tO210curve} becomes
\begin{align}
y^2=4x^3-a_0u^4x+4u^7-b_0u^6.
\label{eq:SWcurve2}
\end{align}
This falls into the singular exception \eqref{eq:singular}
excluded from the moduli space of
$(U\oplus E_8(-1))$-polarized K3 surfaces.
In fact, the equation describes
another Seiberg--Witten curve for the E-string theory
associated with the singular fiber at $u=\infty$.
(This can be seen clearly
in the new coordinates $v=1/u$, $X=x/u^4$, $Y=y/u^6$.)
It is interesting that two independent Seiberg--Witten curves
\eqref{eq:SWcurve1} and \eqref{eq:SWcurve2} arise 
from two different limiting processes
associated with the same limit $p\to 0$.

Second, relations similar to \eqref{eq:tcoeffexp} 
between other orthogonal modular forms
and Weyl invariant Jacobi forms were considered
in \cite{Wang:2020}.
The relations provide a universal way to
prove that the graded rings of orthogonal modular forms
of certain classes are freely generated,
based on the polynomial-ring structure of the associated
bigraded rings of Weyl invariant weak Jacobi forms.
While the proof is not applicable to the present case
because the ring of $W(E_8)$-invariant weak Jacobi forms is not
freely generated,
we have demonstrated that analogous relations \eqref{eq:tcoeffexp}
still exist and play an important role
in the study of lattice-polarized K3 surfaces.

\section{Reductions}\label{sec:red}

In this section we show that our solution reduces to known ones
by adjusting the Wilson line parameters $\vecz$ to special values.
In particular, we see that our result reproduces 
the explicit inverse period map of
\cite{Kumar:2008,Clingher:2012,Malmendier:2014uka}.

We first consider the simplest case
\begin{align}
\vecz=\bvec{0}.
\end{align}
This is the case without Wilson lines
and the heterotic side has an unbroken $E_8\times E_8$ symmetry.
On the F-theory side, we have
algebraic K3 surfaces polarized by
the lattice $S=U\oplus E_8(-1)\oplus E_8(-1)$.
The orthogonal complement is $S^\perp=U\oplus U$
and the corresponding modular forms are
symmetric combinations of elliptic modular forms
in $\tau$ and $\rho$.
Indeed, after some computation we see that
the curve expressed in the form \eqref{eq:tO210curve}
becomes
\begin{align}
y^2=4x^3
 -\frac{E_4(\tau)E_4(\rho)}{12}u^4x
 +4u^7
 -\frac{E_6(\tau)E_6(\rho)}{216}u^6
 +4\Delta(\tau)\Delta(\rho)u^5.
\label{eq:mlcurve2}
\end{align}
This curve can be transformed into the form \eqref{eq:mlcurve1}
by the rescaling
\begin{align}
u\mapsto\eta(\tau)^{12}\eta(\rho)^{12}u,\quad
x\mapsto\eta(\tau)^{28}\eta(\rho)^{28}x,\quad
y\mapsto 2\eta(\tau)^{42}\eta(\rho)^{42}y.
\end{align}
We have thus verified that
our result correctly reproduces the F-theory dual
of the heterotic $E_8\times E_8$ string theory
compactified on $T^2$ without Wilson lines.

Next let us consider a slightly more general case
\begin{align}
\vecz=(0,0,0,0,0,0,z,z).
\label{eq:veczE7}
\end{align}
For a generic value of $z$,
this corresponds to breaking one of the two $E_8$'s down to $E_7$.
On the F-theory side, we have
algebraic K3 surfaces polarized by the lattice
$S=U\oplus E_8(-1)\oplus E_7(-1)$.
The orthogonal complement is $S^\perp=U\oplus U\oplus A_1(-1)$.
Due to the exceptional isomorphism
$\grp{Spin}(2,3)\cong\grp{Sp}(4,\bbR)$,
the corresponding orthogonal modular forms are
identified with Siegel modular forms of genus two.

It is well known that
the graded ring of even-weight Siegel modular forms of genus two
is a polynomial ring,
generated by Igusa's modular forms
$\psi_4,\psi_6,\chi_{10},\chi_{12}$
\cite{Igusa:1962,Igusa:1967}.
In this paper we follow the normalization of
Malmendier and Morrison \cite{Malmendier:2014uka}.
If we adjust the parameters as in \eqref{eq:veczE7},
the orthogonal Eisenstein series $\gen_k$
reduce to even-weight Siegel modular forms of genus two
and hence they are written as polynomials in
$\psi_4,\psi_6,\chi_{10},\chi_{12}$.
Indeed, after some computation we obtain, for instance,
\begin{align}
\gen_4=\frac{\psi_4}{240},\quad
\gen_{10}=-\frac{\psi_4\psi_6}{264}-\frac{12096}{11}\chi_{10},\quad
\gen_{12}=\frac{7}{1040}\psi_4^3+\frac{25}{6552}\psi_6^2
 -\frac{86400}{13}\chi_{12}.
\end{align}
Using these results and the substitution
\begin{align}
u=t-\frac{\psi_6}{6048},
\end{align}
we see that the curve \eqref{eq:O210curve} becomes
\begin{align}
y^2=4x^3-\left(\frac{\psi_4}{12}t^4+16\chi_{10}t^3\right)x
 +4t^7-\frac{\psi_6}{216}t^6+4\chi_{12}t^5.
\end{align}
After replacing $y$ with $2y$, this is in perfect agreement
with the Weierstrass model given in \cite{Malmendier:2014uka}.
It was shown in \cite{Malmendier:2014uka} that
the model is equivalent to the results of Kumar \cite{Kumar:2008}
and Clingher--Doran \cite{Clingher:2012}
that provide an explicit inverse period map for
$(U\oplus E_8(-1)\oplus E_7(-1))$-polarized K3 surfaces.
We have thus shown that our solution reproduces
their inverse period map.
This serves as another evidence for our conjecture.

\section{Conclusions and outlook}\label{sec:conclusion}

We have constructed the precise F-theory backgrounds
dual to the heterotic string theory compactified on $T^2$
with general Wilson lines turned on for one of the two $E_8$'s.
The elliptic curve describing the backgrounds
is expressed in terms of orthogonal modular forms
for the even unimodular lattice of signature $(2,10)$.
We have shown that the curve reduces to both
the Seiberg--Witten curve for the E-string theory
and the curve describing an explicit inverse period map for the
$(U\oplus E_8(-1)\oplus E_7(-1))$-polarized K3 surfaces.
We conjecture that our result gives an explicit inverse period map
for the $(U\oplus E_8(-1))$-polarized K3 surfaces.

There are several possible directions for further study.
Although our solution has been uniquely determined and 
has already passed several nontrivial tests,
it is still important to prove our conjecture.
A natural approach is to compute explicitly the periods 
of the K3 surfaces we have obtained.
This is feasible in principle,
but may require further technical advances.

A more challenging direction would be to generalize
our solution to the full
heterotic/F-theory duality in eight dimensions.
The corresponding general elliptic K3 surfaces with section
are related to modular forms
for the even unimodular lattice of signature $(2,18)$.
The graded ring of these modular forms
is not a polynomial ring \cite{Shvartsman:2017,Wang:2020},
but is isomorphic to the ring of joint invariants of
binary forms of degree $12$ and $8$ 
\cite{Odaka:2018,Nagano:2022}.
This is analogous to the fact that
the ring of $W(E_8)$-invariant weak Jacobi forms
is isomorphic to the ring of joint covariants of
binary forms of degree $6$ and $4$, for which
an explicit isomorphism was established \cite{Sakai:2024vby}.
It would be very interesting to develop a similar understanding
for general elliptic K3 surfaces.


\begin{center}
  {\bf Acknowledgments}
\end{center}

This work was supported in part by JSPS KAKENHI Grant Number 25K07326.


\appendix
\renewcommand{\theequation}{\Alph{section}.\arabic{equation}}

\section{Explicit expressions for coefficient functions}
\label{app:coeff}

%
\begin{align}
\coeff_4&=
  \tnum{20}\gen_4,\qquad
\coeff_{10}=
\hspace{-.2em}{}
 -\tfrac{11}{756}\gen_{10},\qquad
\coeff_{12}=
  \tfrac{13}{21600}\gen_{12}
 -\tnum{56}\gen_4^3,\nn\\[1ex]
\coeff_{16}&=
  \tfrac{11747}{24766560000}\gen_{16}
 -\tfrac{151}{54000}\gen_4\gen_{12}
 +\tfrac{1564}{15}\gen_4^4,\qquad
\coeff_{18}=
\hspace{-.2em}{}
 -\tfrac{19}{2874009600}\gen_{18}
 +\tfrac{29}{216}\gen_4^2\gen_{10},\nn\\[1ex]
\coeff_{22}&=
\hspace{-.2em}{}
 -\tfrac{1008941}{688725936138240000}\gen_{22}
 +\tfrac{223}{5645376000}\gen_4\gen_{18}
 +\tfrac{443}{299376000}\gen_{10}\gen_{12}
 -\tfrac{48157}{124740}\gen_4^3\gen_{10},\nn\\[1ex]
\coeff_{24}&=
  \tfrac{174611}{16690988250832896000}\gen_{24}
 -\tfrac{136127}{26087443200}\gen_4^2\gen_{16}
 -\tfrac{107}{3071520000}\gen_{12}^2
 +\tfrac{21397}{1279800}\gen_4^3\gen_{12}
\nn\\[0ex] &\hspace{1.15em}
 -\tfrac{4385}{45151344}\gen_4\gen_{10}^2
 -\tfrac{632732}{2133}\gen_4^6,\nn\\[1ex]
\coeff_{28}&=
  \tfrac{6854558639}{7896534480915704998133760000}\gen_{28}
 -\tfrac{1070540041}{14317529721564458188800}\gen_4\gen_{24}
 -\tfrac{94723}{4659465071923200}\gen_{10}\gen_{18}
\nn\\[0ex] &\hspace{1.15em}
 -\tfrac{265849121}{6883300674432000000}\gen_{12}\gen_{16}
 +\tfrac{1376710871099}{75525104622240000}\gen_4^3\gen_{16}
 +\tfrac{78149711}{592818717600000}\gen_4\gen_{12}^2
\nn\\[0ex] &\hspace{1.15em}
 -\tfrac{6572030171}{164671866000}\gen_4^4\gen_{12}
 +\tfrac{74652857}{165989240928}\gen_4^2\gen_{10}^2
 +\tfrac{120807735116}{320195295}\gen_4^7,\nn\\[1ex]
\coeff_{30}&=
\hspace{-.2em}{}
 -\tfrac{20395861}{5388594369053184530841600000}\gen_{30}
 +\tfrac{11800223}{613495872344678400}\gen_4^2\gen_{22}
 +\tfrac{1453}{1996861870080000}\gen_{12}\gen_{18}
\nn\\[0ex] &\hspace{1.15em}
 -\tfrac{693841}{2288070892800}\gen_4^3\gen_{18}
 +\tfrac{1215469}{173454683351040}\gen_4\gen_{10}\gen_{16}
 -\tfrac{844211}{26000805600}\gen_4^2\gen_{10}\gen_{12}
\nn\\[0ex] &\hspace{1.15em}
 +\tfrac{21289}{1000700096256}\gen_{10}^3
 +\tfrac{28976617}{14444892}\gen_4^5\gen_{10},\nn\\[1ex]
\coeff_{36}&=
  \tfrac{7709321041217}{18413630715760061050043280182476800000000}
  \gen_{36}
 -\tfrac{7764407461783123}{575809653966601125817486540800000}
  \gen_4^2\gen_{28}
\nn\\[0ex] &\hspace{1.15em}
 -\tfrac{3808440521}{3608328531309411926016000000}
  \gen_{12}\gen_{24}
 +\tfrac{12805116916131707}{18270420708566885539000320000}
  \gen_4^3\gen_{24}
\nn\\[0ex] &\hspace{1.15em}
 -\tfrac{25884540899279}{1063086665395272944025600000}
  \gen_4\gen_{10}\gen_{22}
 -\tfrac{1445131}{351346497693224140800000}
  \gen_{18}^2
\nn\\[0ex] &\hspace{1.15em}
 +\tfrac{49307301599077}{87206315312861015040000}
  \gen_4^2\gen_{10}\gen_{18}
 -\tfrac{151224258987701}{1252367573776510464000000000}
  \gen_4\gen_{16}^2
\nn\\[0ex] &\hspace{1.15em}
 +\tfrac{188811616415800243}{173477955500338834800000000}
  \gen_4^2\gen_{12}\gen_{16}
 -\tfrac{15644931691}{7225197541018329600000}
  \gen_{10}^2\gen_{16}
\nn\\[0ex] &\hspace{1.15em}
 -\tfrac{23082211106900137133}{192753283889265372000000}
  \gen_4^5\gen_{16}
 +\tfrac{42161803}{45152965762560000000}
  \gen_{12}^3
\nn\\[0ex] &\hspace{1.15em}
 -\tfrac{11855784862206377}{6051908442363120000000}
  \gen_4^3\gen_{12}^2
 +\tfrac{417274116709}{21903703691417856000}
  \gen_4\gen_{10}^2\gen_{12}
 +\tfrac{10861489642732237}{44239096800900000}
  \gen_4^6\gen_{12}
\nn\\[0ex] &\hspace{1.15em}
 -\tfrac{16307485486218397}{3624420722704135200}
  \gen_4^4\gen_{10}^2
 -\tfrac{234515939414512448}{116742061002375}
  \gen_4^9,\nn\\[1ex]
\coeff_{42}&=
\hspace{-.2em}{}
 -\tfrac{6631038857641423}
       {385139089599291460361143145057199808787251200000000}
  \gen_{42}
\nn\\[0ex] &\hspace{1.15em}
 +\tfrac{3439675631181421}{10042236578185938931213366622945280000000}
  \gen_{12}\gen_{30}
\nn\\[0ex] &\hspace{1.15em}
 +\tfrac{101793602990327293}{464918360101200876445063269580800000}
  \gen_4^3\gen_{30}
\nn\\[0ex] &\hspace{1.15em}
 +\tfrac{57367014447815444341}{3531853925865639809789273061111889920000}
  \gen_4\gen_{10}\gen_{28}
\nn\\[0ex] &\hspace{1.15em}
 +\tfrac{525557356787009}{41387716941670508678931554500608000000}
  \gen_{18}\gen_{24}
\nn\\[0ex] &\hspace{1.15em}
 -\tfrac{127865777066374524367}{224131209549021226924163430875136000}
  \gen_4^2\gen_{10}\gen_{24}
\nn\\[0ex] &\hspace{1.15em}
 +\tfrac{57308022955055820763}{70640630731886742500551203225600000000}
  \gen_4\gen_{16}\gen_{22}
\nn\\[0ex] &\hspace{1.15em}
 -\tfrac{53188340809834246477}{29726237858133048366052638720000000}
  \gen_4^2\gen_{12}\gen_{22}
 +\tfrac{2128537307765681}{296394254819105213289026027520000}
  \gen_{10}^2\gen_{22}
\nn\\[0ex] &\hspace{1.15em}
 -\tfrac{14896552215264461009}{18431447084655907965062400000}
  \gen_4^5\gen_{22}
 -\tfrac{2827301790511484477}{320203543876808129386168320000000}
  \gen_4^2\gen_{16}\gen_{18}
\nn\\[0ex] &\hspace{1.15em}
 -\tfrac{31649332753163}{1175953265177184129613824000000}
  \gen_{12}^2\gen_{18}
 +\tfrac{531162017398011833}{134744644968219014851584000000}
  \gen_4^3\gen_{12}\gen_{18}
\nn\\[0ex] &\hspace{1.15em}
 -\tfrac{78228771812432399}{240096678267068193520502784000}
  \gen_4\gen_{10}^2\gen_{18}
 +\tfrac{191897548050681874939}{18714534023363752062720000}
  \gen_4^6\gen_{18}
\nn\\[0ex] &\hspace{1.15em}
 +\tfrac{90583520754707467}{1280278213931541657614745600000000}
  \gen_{10}\gen_{16}^2
\nn\\[0ex] &\hspace{1.15em}
 -\tfrac{19370095325355922553657}{16340033429622263043660257280000000}
  \gen_4\gen_{10}\gen_{12}\gen_{16}
\nn\\[0ex] &\hspace{1.15em}
 -\tfrac{141102328089868242818386937}{627502672679243851607230713600000}
  \gen_4^4\gen_{10}\gen_{16}
 +\tfrac{1726659384576090512671}{788072430366397301861514240000}
  \gen_4^2\gen_{10}\gen_{12}^2
\nn\\[0ex] &\hspace{1.15em}
 -\tfrac{253320452817037}{74596575078126898900992000}
  \gen_{10}^3\gen_{12}
 +\tfrac{28241189020977720247709}{35537176694011422342240000}
  \gen_4^5\gen_{10}\gen_{12}
\nn\\[0ex] &\hspace{1.15em}
 +\tfrac{1618112052716582733563}{1755252231270612172327918080}
  \gen_4^3\gen_{10}^3
 -\tfrac{2554914109707652713065783}{54292908838073006356200}
  \gen_4^8\gen_{10}.
\label{eq:coeffs}
\end{align}
Here, $\gen_k$ are the Eisenstein series \eqref{eq:genFJ}.
The above expressions are also provided
in the ancillary file \texttt{coeffs.txt}.


\renewcommand{\section}{\subsection}
\renewcommand{\refname}

{\bf References}

{\footnotesize
\begin{thebibliography}{100}

\bibitem{Vafa:1996xn}
C.~Vafa,
``Evidence for F-Theory,''
Nucl. Phys. B \textbf{469} (1996), 403--418
[arXiv:hep-th/9602022 [hep-th]].

\bibitem{Morrison:1996na}
D.~R.~Morrison and C.~Vafa,
``Compactifications of F-Theory on Calabi--Yau threefolds (I),''
Nucl. Phys. B \textbf{473} (1996), 74--92
[arXiv:hep-th/9602114 [hep-th]].

\bibitem{Morrison:1996pp}
D.~R.~Morrison and C.~Vafa,
``Compactifications of F-Theory on Calabi--Yau threefolds (II),''
Nucl. Phys. B \textbf{476} (1996), 437--469
[arXiv:hep-th/9603161 [hep-th]].

\bibitem{Inose:1977}
H.~Inose,
``Defining equations of singular K3 surfaces and a notion of isogeny,''
in \textit{Proceedings of the International Symposium on Algebraic
Geometry (Kyoto Univ., Kyoto, 1977)},
Kinokuniya Book Store, Tokyo (1978), 495--502.

\bibitem{Kumar:2008}
A.~Kumar,
``K3 surfaces associated with curves of genus two,''
Int. Math. Res. Not. IMRN \textbf{2008} (2008), rnm165
[arXiv:math/0701669 [math.AG]].

\bibitem{Clingher:2012}
A.~Clingher and C.~F.~Doran,
``Lattice polarized K3 surfaces and Siegel modular forms,''
Adv. Math. \textbf{231}(1) (2012), 172--212
[arXiv:1004.3503 [math.AG]].

\bibitem{Malmendier:2014uka}
A.~Malmendier and D.~R.~Morrison,
``K3 surfaces, modular forms, and non-geometric heterotic
compactifications,''
Lett. Math. Phys. \textbf{105}(8) (2015), 1085--1118
[arXiv:1406.4873 [hep-th]].

\bibitem{Nagano:2021}
A.~Nagano,
``Inverse period mappings of K3 surfaces and a construction of modular
forms for a lattice with the Kneser conditions,''
J. Algebra \textbf{565} (2021), 33--63
[arXiv:1903.01282 [math.AG]].

\bibitem{Clingher:2025}
A.~Clingher, A.~Malmendier and B.~Williams,
``K3 Surfaces and Orthogonal Modular Forms,''
Nagoya Math. J. \textbf{260} (2025), 687--727
[arXiv:2411.05970 [math.AG]].

\bibitem{Clingher:2026}
A.~Clingher, A.~Malmendier and B.~Williams,
``On the Vinberg Family of K3 Surfaces,''
[arXiv:2606.07849 [math.AG]].

\bibitem{Hashimoto:2022}
K.~Hashimoto and K.~Ueda,
``The ring of modular forms for the even unimodular lattice of signature
(2,10),''
Proc. Amer. Math. Soc. \textbf{150}(2) (2022), 547--558
[arXiv:1406.0332 [math.AG]].

\bibitem{Looijenga:1984}
E.~Looijenga,
``The smoothing components of a triangle singularity. II,''
Math. Ann. \textbf{269}(3) (1984), 357--387.

\bibitem{Dieckmann:2019}
C.~Dieckmann, A.~Krieg, and M.~Woitalla,
``The graded ring of modular forms on the Cayley half-space of degree
two,''
Ramanujan J. \textbf{48}(2) (2019), 385--398
[arXiv:1707.05029 [math.NT]].

\bibitem{Ganor:1996mu}
O.~J.~Ganor and A.~Hanany,
``Small $E_8$ Instantons and Tensionless Non Critical Strings,''
Nucl. Phys. B \textbf{474} (1996), 122--140
[arXiv:hep-th/9602120 [hep-th]].

\bibitem{Seiberg:1996vs}
N.~Seiberg and E.~Witten,
``Comments on String Dynamics in Six Dimensions,''
Nucl. Phys. B \textbf{471} (1996), 121--134
[arXiv:hep-th/9603003 [hep-th]].

\bibitem{Ganor:1996pc}
O.~J.~Ganor, D.~R.~Morrison and N.~Seiberg,
``Branes, Calabi--Yau Spaces, and Toroidal Compactification of
the $N$=1 Six-Dimensional $E_8$ Theory,''
Nucl. Phys. B \textbf{487} (1997), 93--127
[arXiv:hep-th/9610251 [hep-th]].

\bibitem{Eguchi:2002fc}
T.~Eguchi and K.~Sakai,
``Seiberg--Witten Curve for the $E$-String Theory,''
JHEP \textbf{05} (2002), 058
[arXiv:hep-th/0203025 [hep-th]].

\bibitem{Minahan:1998vr}
J.~A.~Minahan, D.~Nemeschansky, C.~Vafa and N.~P.~Warner,
``$E$-Strings and $N=4$ Topological Yang-Mills Theories,''
Nucl. Phys. B \textbf{527} (1998), 581--623
[arXiv:hep-th/9802168 [hep-th]].

\bibitem{Eguchi:2002nx}
T.~Eguchi and K.~Sakai,
``Seiberg--Witten Curve for $E$-String Theory Revisited,''
Adv. Theor. Math. Phys. \textbf{7}(3) (2003), 419--455
[arXiv:hep-th/0211213 [hep-th]].

\bibitem{Sakai:2012ik}
K.~Sakai,
``Seiberg--Witten prepotential for E-string theory and global
symmetries,''
JHEP \textbf{09} (2012), 077
[arXiv:1207.5739 [hep-th]].

\bibitem{Sakai:2017ihc}
K.~Sakai,
``$E_n$ Jacobi forms and Seiberg--Witten curves,''
Commun. Num. Theor. Phys. \textbf{13} (2019), 53--80
[arXiv:1706.04619 [hep-th]].

\bibitem{Narain:1985jj}
K.~S.~Narain,
``New Heterotic String Theories in Uncompactified Dimensions
{\ensuremath{<}} 10,''
Phys. Lett. B \textbf{169} (1986), 41--46.

\bibitem{Narain:1986am}
K.~S.~Narain, M.~H.~Sarmadi and E.~Witten,
``A Note on Toroidal Compactification of Heterotic String Theory,''
Nucl. Phys. B \textbf{279} (1987), 369--379.

\bibitem{Clingher:2003ui}
A.~Clingher and J.~W.~Morgan,
``Mathematics underlying the F theory / Heterotic string duality
in eight-dimensions,''
Commun. Math. Phys. \textbf{254} (2005), 513--563
[arXiv:math/0308106 [math.AG]].

\bibitem{Nikulin:1979}
V.~V.~Nikulin,
``Finite groups of automorphisms of K\"ahlerian~$K_3$ surfaces,''
Tr. Mosk. Mat. Obs. \textbf{38} (1979), 75--137.

\bibitem{Dolgachev:1996xw}
I.~V.~Dolgachev,
``Mirror symmetry for lattice polarized K3 surfaces,''
J. Math. Sci. \textbf{81} (1996), 2599--2630
[arXiv:alg-geom/9502005 [math.AG]].

\bibitem{Pyatetski-Shapiro:1971}
I.~I.~Pjatecki\u{\i}-\v{S}apiro and I.~R.~\v{S}afarevi\v{c},
``A Torelli theorem for algebraic surfaces of type K3,''
Math. USSR Izv. \textbf{5}(3) (1971), 547--588.

\bibitem{Burns:1975}
D.~Burns and M.~Rapoport,
``On the Torelli problem for k\"{a}hlerian $K-3$ surfaces,''
Ann. Sci. \'{E}c. Norm. Sup\'{e}r. \textbf{8}(2) (1975), 235--273.

\bibitem{Todorov:1980}
A.~N.~Todorov,
``Applications of the K\"{a}hler-Einstein-Calabi-Yau metric to moduli
of K3 surfaces,''
Invent. Math. \textbf{61}(3) (1980), 251--265.

\bibitem{Wang:2018fil}
H.~Wang,
``Weyl invariant $E_8$ Jacobi forms,''
Commun. Num. Theor. Phys. \textbf{15}(3) (2021), 517--573
[arXiv:1801.08462 [math.NT]].

\bibitem{Sakai:2024vby}
K.~Sakai,
``The ring of Weyl invariant $E_8$ Jacobi forms,''
Math. Z. \textbf{313}(4) (2026), 71
[arXiv:2410.12907 [math.NT]].

\bibitem{Eichler:1985}
M.~Eichler and D.~Zagier,
``The Theory of Jacobi forms,''
Prog. in Math. \textbf{55}, Birkh\"auser-Verlag, 1985.

\bibitem{Wirthmuller:1992}
K.~Wirthm\"uller,
``Root systems and Jacobi forms,''
Compos. Math. \textbf{82} (1992), 293--354.

\bibitem{Sakai:2022taq}
K.~Sakai,
``Algebraic construction of Weyl invariant $E_8$ Jacobi forms,''
J. Number Theory \textbf{244} (2023), 42--62
[arXiv:2201.06895 [math.NT]].

\bibitem{Sakai:2011xg}
K.~Sakai,
``Topological string amplitudes for the local $\frac{1}{2}$K3
surface,''
PTEP \textbf{2017}(3) (2017), 033B09
[arXiv:1111.3967 [hep-th]].

\bibitem{Gritsenko:2018}
V.~A.~Gritsenko,
``Reflective modular forms and applications,''
Russian Math. Surveys \textbf{73}(5) (2018), 797--864.

\bibitem{Eie:1991}
M.~Eie,
``The Maa\ss{} space for Cayley numbers,''
Math. Z. \textbf{207}(1) (1991), 645--655.

\bibitem{Wang:2020}
H.~Wang, and B.~Williams,
``On some free algebras of orthogonal modular forms,''
Adv. Math. \textbf{373} (2020), 107332
[arXiv:2003.05374 [math.NT]].

\bibitem{Igusa:1962}
J.~Igusa,
``On Siegel modular forms of genus two,''
Amer. J. Math. \textbf{84}(1) (1962), 175--200;
(II), ibid.~\textbf{86}(2) (1964), 392--412.

\bibitem{Igusa:1967}
J.~Igusa,
``Modular forms and projective invariants,''
Amer. J. Math. \textbf{89}(3) (1967), 817--855.

\bibitem{Shvartsman:2017}
O.~V.~Shvartsman and E.~B.~Vinberg,
``A criterion of smoothness at infinity for an arithmetic quotient of
the future tube,''
Funct. Anal. Its Appl. \textbf{51} (2017), 32--47.

\bibitem{Nagano:2022}
A.~Nagano and K.~Ueda,
``The ring of modular forms for the even unimodular lattice of
signature (2,18),''
Hiroshima Math. J. \textbf{52}(1) (2022), 43--51
[arXiv:2102.09224 [math.AG]].

\bibitem{Odaka:2018}
Y.~Odaka and Y.~Oshima,
``Collapsing K3 Surfaces, Tropical Geometry and Moduli Compactifications
of Satake, Morgan-Shalen Type,''
MSJ Mem. \textbf{40} (2021)
[arXiv:1810.07685 [math.AG]].

\end{thebibliography}
}
\end{document}